\documentclass[aps,prc,reprint,superscriptaddress,floatfix,onecolumn,nofootinbib]{revtex4-2}
\usepackage{epsfig,rotating,graphics}
\usepackage{amsmath}
\usepackage{amsfonts}
\usepackage{amssymb}
\usepackage{amsthm}
\usepackage{caption}
\usepackage{fontenc}
\usepackage{graphicx}
\usepackage{blindtext}
\usepackage{comment}
\usepackage{tikz}
\usepackage{soul}
\usepackage[normalem]{ulem}
\def\be{\begin{eqnarray} &&} 
 
\def\ee{\end{eqnarray}}

\newcommand{\bfi}{\begin{figure}}
\newcommand{\efi}{\end{figure}}

\newcommand{\blue}[1]{{\textcolor[rgb]{0,0,1}{#1}}}

\newcommand{\vicente}[1]{{ \blue{{\bf Vicente}:\ {#1}}}}

\begin{document}

\title{Hybrid spectroscopy within the Graviton Soft-Wall model}

\author{Matteo Rinaldi}
\email[Corresponding Author: ]{matteo.rinaldi@pg.infn.it}

\affiliation{Istituto Nazionale di Fisica Nucleare
Sezione di Perugia, Via A. Pascoli, Perugia, Italy}

\author{Vicente Vento}

\affiliation{Departamento de F\'{\i}sica Te\'orica-IFIC, Universidad de Valencia- CSIC,
46100 Burjassot (Valencia), Spain.}

\date{\today}

\begin{abstract}
    {In this analysis, the so called holographic graviton soft-wall model (GSW), first developed to investigate the glueball spectrum, has been adopted to predict the masses of  hybrids with different quantum numbers. Results have been compared with other  models and lattice calculations.
    We have extended the GSW model by introducing two modifications based on anomalous dimensions in order to improve our agreement with other calculations and to remove the initial degeneracy not accounted for by lattice predictions. These modifications do not involve new parameters. The next step has been to identify which of our calculated states agree with the PDG data, leading to experimental hybrids. The procedure has been extended to include hybrids made of heavy quarks by incorporating the quark masses into the model.}
\end{abstract}

\maketitle

\section{Introduction}

In the last few years, hadronic models, inspired by  the holographic conjecture 
\cite{Maldacena:1997re,Witten:1998zw}, have been vastly used and developed in order 
 to investigate 
non-perturbative features of glueballs and mesons, thus trying to 
grasp fundamental features  of QCD 
\cite{Fritzsch:1973pi,Fritzsch:1975wn}. Recently we have 
used the so 
called 
AdS/QCD models to study the scalar glueball 
spectrum~\cite{Vento:2017ice,Rinaldi:2017wdn}. The holographic principle relies in a correspondence between a 
five dimensional 
classical theory with an AdS metric and a supersymmetric
 conformal quantum 
field theory with $N_C  \rightarrow \infty$. This theory, different from QCD, is taken as a starting point  to construct a 5 dimensional 
holographic dual of it. This is the so called bottom-up approach~\cite{Polchinski:2000uf,Brodsky:2003px,DaRold:2005mxj,Karch:2006pv}. 
In this scenario,  models are constructed by modifying  the five dimensional  
classical AdS theory with the aim of resembling  QCD as much as 
possible. The main differences characterizing these models are related to the 
strategy used to break conformal invariance.
Moreover, it must be noted that 
the relation which these models establish with QCD is at the level of the 
leading order in the number of colours expansion, and thus the mesonic and 
glueball spectrum and their decay properties are ideal observables to be studied 
by these models.
The starting point for 
the present investigation is  the holographic
Soft-Wall (SW) model scheme, were a dilaton field  
is introduced to softly break conformal invariance. 
Within this scheme we have recently  introduced 
 the graviton soft-wall model 
(GSW)~\cite{Rinaldi:2017wdn,Rinaldi:2018yhf,Rinaldi:2020ssz} which
has been able to reproduce, not only the scalar meson spectrum, but also the 
lattice QCD scalar glueball 
masses~\cite{Morningstar:1999rf,Chen:2005mg,Lucini:2004my}, 
that was not described by the traditional SW models.  Moreover,  a formalism to 
study the glueball-meson mixing conditions {has been developed}  and  some 
predictions, regarding the observably of pure glueball 
states, {has been provided} ~\cite{Rinaldi:2018yhf,Rinaldi:2020ssz}.
 The success of the model  in reproducing the scalar QCD spectra, has motivated 
us to 
 extend the  GSW model to describe the spectrum of the  $\rho$ vector meson, the $a_1$ axial vector meson, the 
pseudo-scalar meson spectra and high
spin glueballs~\cite{Rinaldi:2021dxh}. 

For forty years of the study of Quantum Chromodynamics (QCD) has served to establish one of the pillars of the Standard Model. 
Although gluons are now firmly established as the carriers of the strong force, their nonperturbative behavior remains
enigmatic. This unfortunate circumstance is chiefly due to two features of QCD: the theory is notoriously difficult to
work with in the nonperturbative regime, and experimental manifestations of glue tend to be hidden in the spectrum
and dynamics of the conventional hadrons.
In particular, experimental
manifestations of
hadrons that carry valence quark and gluonic degrees of freedom have been postulated since the early days of QCD.
 These states are called hybrids, and our lies in extending our 
previous experience with conventional hadrons to these states using the very successful GSW model \cite{Rinaldi:2021dxh}.

Let us describe briefly the contents of this work. In Section~\ref{gsw} 
we summarize the essence of the GSW model 
~\cite{Rinaldi:2021dxh}.
In Section~\ref{lighthybrid} we apply the GSW model to calculate the
spectrum of hybrid states. 
In Section~\ref{lightlattice} we compare our results with lattice QCD and model calculations. 
{In Section \ref{modi}, we present two possible modifications of the GSW model which do not involve new free parameters and allow to reproduce the essential outcome of lattice data.}
In Section~\ref{lightpheno} we compare our results with experimental states appearing in the PDG compilation with the same quantum numbers. From Section~\ref{heavyhybrid} to Section~\ref{heavypheno}, we repeat the  analysis for the heavy particles, i.e., we discuss the  GSW model predictions for the spectra, we compare our results with lattice QCD, model calculations and experimental data. We end by collecting some conclusions of our study.

\section{Description of Hadrons in the GSW model}
\label{gsw}
In this section, the essential features  of the GSW model are introduced. The development of this approach has been motivated by the impossibility of the conventional SW models to  describe
the glueball and meson spectra with the same energy 
scale~\cite{Rinaldi:2017wdn,Rinaldi:2018yhf,Rinaldi:2020ssz}. The main differences, which distinguishes the GSW model from the traditional SW, is a 
deformation of the $AdS$ metric in 5 dimensions.

\begin{equation}
ds^2=\frac{R^2}{z^2} e^{\alpha \phi_0(z)} (dz^2 + \eta_{\mu \nu} 
dx^\mu dx^\nu) = e^{2 A(z)}  (dz^2 + \eta_{\mu \nu} 
dx^\mu dx^\nu) = e^{\alpha {\phi_0(z)}} g_{M N} dx^M 
dx^N  =\bar{g}_{MN}dx^M dx^N.
\label{metric5}
\end{equation}
where $A(z) = \log{{R}/{z} }+{\alpha \phi_0(z)}/{2}$.

The quantities evaluated within this new metric are displayed with overline. The 
function $\phi_0(z)$ will be specified later. This kind of 
modification has been adopted in many studies of the properties of mesons 
and glueballs within AdS/QCD 
~\cite{Colangelo:2007pt,Capossoli:2015ywa,Vega:2016gip,Akutagawa:2020yeo,
Gutsche:2019blp, Klebanov:2004ya,MartinContreras:2019kah,FolcoCapossoli:2019imm,
Bernardini:2016qit,Li:2013oda}. 
{The metric tensor and its determinant of this new space can be related to the usual $AdS_5$ metric and its determinant:}

\begin{eqnarray}
\bar g^{MN} &= &e^{-\alpha \phi_0(z)} g^{MN}, \\
\sqrt{-\bar{g} } & = & e^{\frac{5}{2}\alpha \phi_0(z)} \sqrt{-g}.
\end{eqnarray}

Once the gravitational background has been defined by the model, the same 
strategy used in the SW case is considered in order to obtain the equations of 
motion (EOMs) for the different fields dual to given hadronic states. The new action, {written} in terms of the standard AdS metric of the  SW model,
is given by

\begin{align}
\label{prefactor}
 \bar S ={\int d^4xdz~e^{-\phi_0(z) \beta  }\sqrt{-\bar g} 
\mathcal{L}(x_\mu,z)}= \int d^4xdz~e^{\phi_0(z) 
\left(\frac{5}{2} \alpha -\beta+1  
\right) }\sqrt{-g} e^{-\phi_0(z)} \mathcal{L}(x_\mu,z)~,
\end{align}
 where here the prefactor $exp\left[ {\phi_0(z) \left(\frac{5}{2} \alpha 
+\beta+1\right)} \right]$ {takes into account the dilaton term, as in the SW model, and also}  the modification of the metric.   {The 
parameters $\alpha$ and  $\beta$ parametrise the internal 
dynamics of the hadrons of QCD {described within this holographic framework}. In the AdS dynamics}, $\alpha$ characterises the 
modification of the metric, while $\beta$  characterises the SW model dilaton, 
namely the breaking of conformal invariance.
{Since the GSW model has been developed as a modification of the SW model, we propose to fix $\beta$ to reproduce the kinetic term of the standard SW model action} ~\cite{Rinaldi:2017wdn,Rinaldi:2018yhf,Rinaldi:2020ssz,Rinaldi:2021dxh}. { Thus 
in the case of scalar fields,  $\beta= \beta_s= 1+\frac{3}{2} \alpha$ 
and in the case of the vector { fields}  $\beta=\beta_\rho=
1+\frac{1}{2}\alpha$.} The 
function $\mathcal{L}(x_\mu,z)$ is the Lagrangian density {describing the motion of the dual fields in the space described by the metric Eq. (\ref{metric5})}. 
{The dilaton profile function $\phi_0$ adopted in the GSW model is the same of the one usually addressed in SW based models \cite{Erlich:2005qh,Colangelo:2008us,deTeramond:2005su,Rinaldi:2017wdn,
Colangelo:2007pt,Capossoli:2015ywa,BoschiFilho:2005yh,FolcoCapossoli:2019imm}, i.e.$\phi_0(z)= k^2 z^2$  }.
The action characterizing the fields propagating in the $AdS_5$ space contains a mass like term whose value is fixed as follows:

\begin{equation}
M_5^2R^2 = (\Delta +  - p)(\Delta  +p -4).
\label{M5A}
\end{equation}
where $\Delta$ is the conformal dimension of the fields and $p$ depends on its  $p$-form. {In other analyses, $\Delta$ has been corrected by including the contribution of an }   the anomalous conformal dimension that
characterises  the  chiral symmetry breaking mechanism \cite{Contreras:2018hbi}. In particular, 
$\Delta_p=0$ for scalar, {vector and tensor mesons,} and $\Delta_p=-1$ for pseudoscalar and axial vector mesons. 
{Here and in the next sections, we the GSW predictions for the hybrid masses without taking into account further modifications of the model.}

Within these scheme, hybrids and multiquark states, 
{defined as quark and gluon operators in QCD}, will be described by the properties of their $p$-forms. 
For example, a vector field can be described as~\cite{Rinaldi:2021dxh}

\begin{equation}
V_\mu=\bar{\Psi} \gamma_\mu\Psi
\end{equation}
thus $p=1$, $\Delta=3$ and therefore $M_5^2R^2 = 0$.
{Within the present prescription, an }  
 an axial vector field

\begin{equation}
A_\mu =\bar{\Psi} \gamma_\mu \gamma_5\Psi
\end{equation}
has   $M_5^2R^2 = 0$. In the next section, we extend the calculation of the $AdS_5$ mass to hybrids.

\section{Description of the light hybrids and their spectrum}
 \label{lighthybrid}

Hybrids are hadrons formed of valence quarks and valence gluons. Due to the non perturbative nature of this bound states, the use of models is needed to predict and describe possible properties of these systems, see e.g. Refs. \cite{Chanowitz:1982qj,Barnes:1982tx,Isgur:1984bm,Barnes:1995hc,Ishida:1991mx,General:2006ed}.
In this scenario, the GSW model, already successfully applied to the studies of glueballs and regular mesons, can be used to calculate the spectra of hybrids with different quantum numbers. In particular, let us start with 
only non strange light hybrids described as quark-antiquark color octet coupled to a valence gluon leading to a hadronic color singlet.

Let us describe the {hybrid} fields with the lowest conformal dimensions and characterized by the following  quantum numbers  $J^{P C}$, i.e. spin, parity and charge conjugation. Since these hybrid fields have to be gauge invariant,  we will use only gauge invariant quantities for the quarks and gluon counterparts. In order to construct the fields, we will use for the quarks  conventional bilinears and for the gluons their color magnetic field $B_i^a = -\frac{1}{2}\varepsilon^{ijk}  F_{ij}^a$, where $F_{\mu \nu}^a$ is the color gauge tensor, $\mu,\nu =0,1,2,3$ are Lorentz indices, $i,j,k=1,2,3$ are the spatial Lorentz indices and $a$ is the color index, and  the color electric field $E_i^a=F_{0i}^a$. To describe the quantum numbers of the hybrids, we recall the transformation properties of the quark bilinears and color magnetic and electric  fields under $C$ and $P$,

\begin{align}
\mbox{scalar}\;\; & \bar{\Psi} ^P \Psi^P = {\Psi} \Psi \; ;\qquad \qquad \quad \quad\bar{\Psi}^C \Psi^C = {\Psi}\Psi \nonumber \\
\mbox{pseudoscalar} \;\;& \bar{\Psi}^P\gamma_5 \Psi^P =  - \bar{\Psi}\gamma_5 \Psi \; ;\qquad\; \quad\bar{\Psi}^C\gamma_5 \Psi^C = \bar{\Psi}\gamma_5 \Psi \nonumber \\
\mbox{vector} \;\; &\bar{\Psi}^P\gamma_\mu \Psi^P = - \bar{\Psi}\gamma_\mu \Psi \; ; \quad \; \qquad  \bar{\Psi}^C\gamma_\mu \Psi^C =- \bar {\Psi}\gamma_\mu \Psi \nonumber \\
\mbox{axial vector} \;\; &\bar{\Psi}^P\gamma_5\gamma_\mu \Psi^P = - \bar{\Psi}\gamma_5\gamma_\mu \Psi \; ; \; \quad \bar{\Psi}^C\gamma_5\gamma_\mu \Psi^C = \bar{\Psi}\gamma_5\gamma_\mu \Psi \nonumber \\
\mbox{tensor} \;\;& \bar{\Psi}^P\sigma_{\mu\nu} \Psi^P =  {\Psi}\sigma_{\mu\nu} \Psi \; ; \qquad \quad  \bar{\Psi}^C\sigma_{\mu\nu}\Psi^C = - \bar{\Psi}\sigma_{\mu \nu} \Psi \nonumber \\
\mbox{magnetic field} \;\;& P^+ B_i^a P = B_i^a \; ; \qquad \qquad \qquad \! \!C^+ B_i^a C = - B_i^a\nonumber \\
\mbox{electric field}\;\;\; &P^+ E_i^a P = - E_i^a \; ;  \qquad \qquad \quad C^+ E_i^a C = - E_i^a \nonumber \\
\label{JPC}
\end{align}

By using these properties, {one can build hybrid field configurations for specific quantum numbers. A crucial role in the solution of the dual field  equation of motion is played by the $AdS_5$ mass Eq.(\ref{M5A}), which strongly depends  on the value of the p-forms. In order to calculate the hybrid ground states, } we  consider the configurations corresponding to the minimum $AdS_5$ mass. Taking these arguments into account, the hybrid field configurations with the lowest $AdS_5$ masses can be divided in two classes. The first one corresponds to those which have mesonic quantum numbers, and the corresponding hadrons are less wishful phenomenologically. 
These fields, to lowest order in conformal dimensions, are

 \begin{align}
 0^{- +} \; \;\;\;\;\,\;S& =  \bar{\Psi} \gamma^i\lambda^a \Psi B_i^a \qquad \qquad \;(\Delta = 5, p=0, M_5^2R^2=\;5), \nonumber \\
 0^{+ +} \; \;\;\;\:\:S^{\prime} &=  \varepsilon_{i j k} \bar{\Psi}  \sigma^{i j} \lambda^a \Psi  B^{k a} \qquad (\Delta = 5, p=0, M_5^2R^2=\;5), \nonumber \\
 1^{- -} \; \;\;\, \;\;V_i&=  \bar{\Psi} \gamma_5  \lambda^a \Psi B_i^a \qquad \qquad \;(\Delta = 5 , p=1, M_5^2R^2=\;8),\nonumber \\
  1^{+-}\;\;\;\;\; A^\prime_i &=\varepsilon_{i j k} \bar{\Psi} \gamma_5\gamma^j \lambda^a \Psi B^{k a}\; \quad (\Delta = 5, p=1, M_5^2R^2=\; 8), \nonumber \\
 1^{+ +} \; \;\;\;\;A_i &=  \varepsilon_{i j k} \bar{\Psi} \gamma^j \lambda^a \Psi E^{k a} \qquad \;(\Delta = 5, p=1, M_5^2R^2=\; \,8). \nonumber \\
\label{mesonicfields}
  \end{align}
The second class are those field configurations with similar properties but with exotic quantum numbers, i.e. quantum numbers that cannot be obtained by mesonic quark-antiquark states. To lowest order in conformal dimensions, these fields are

\begin{align}
 0^{+ -} \; \;\;\;\;\,\, \;\Sigma& =  \bar{\Psi} \gamma_5 \gamma^i \lambda^a \Psi B_i^a \qquad \qquad (\Delta = 5, p=0, M_5^2R^2=\;5), \nonumber \\
 0^{- -} \; \;\;\,\, \;\;\Sigma^\prime&=  \bar{\Psi} \gamma_5  \lambda^a \Psi E_i^a \qquad \qquad  \quad (\Delta = 5 , p=0, M_5^2R^2=\;5),\nonumber \\
 1^{- +} \; \;\;\;\;W_i &=  \varepsilon_{i j k} \bar{\Psi} \gamma^j \lambda^a \Psi B^{k a} \qquad \quad (\Delta = 5, p=1, M_5^2R^2=\;8), \nonumber \\
 1^{- +} \; \, \;\; \, \,W_i^{\prime} &=  \bar{\Psi} \gamma_0\lambda^a \Psi  E_i^a \qquad \qquad \quad (\Delta = 5, p=1, M_5^2R^2=\;\,8). \nonumber \\
\label{exoticfields}
\end{align}

In {the present analysis, we  }
only study $J=0$ and $J=1$ light hybrids.
{The equation of motion for the dual fields of these states are essentially those corresponding to}
 scalar and vector fields, respectively. {As shown in Ref. \cite{Rinaldi:2021dxh}, the potential term strongly depends on} the $AdS_5$ masses given in Eqs. (\ref{mesonicfields}) and (\ref{exoticfields}). Let us point out that the EOMs  {can be re-arranged as a Schr\"odinger like equation.}
For the scalars, the EOMs reads:
\begin{equation}
 -\frac{d^2 \sigma(z)}{dz^2} + 
\left( k^4z^2 + 2 k^2 + \frac{15}{4z^2} 
+M_5^2R^2 \frac{e^{\alpha k^2 z^2}}{z^2}  \right) \sigma(z) = M^2 \sigma(z).
\label{scalar}
\end{equation}
and the EOMs for the vectors are

\begin{align}
 -\psi''(z)+ \left(k^4 z^2 +\frac{3}{4 z^2}+  M_5^2R^2 \frac{e^{\alpha k^2 z^2} }{z^2}  
\right) \psi(z) =M^2 \psi(z)~.
\label{vector}
\end{align}
where $M^2$ is related to the mode energies. We must recall that the parameters of the model have been fixed in our previous works,  being $\alpha=0.55\pm 0.04$ and $k= \frac{370}{\sqrt{\alpha}}$ MeV \cite{Rinaldi:2021dxh}. {For the sake of simplicity, we display results corresponding to $\alpha= 0.55$.}

\begin{table} [htb]
\begin{center}
 mesonic  hybrids \\
\begin{tabular} {|c |c| c| c |c |c |c|}
\hline
 & n=0 & n=1 & n=2 & n=3 \\ \hline
$0^{-+}$ &2074& 2536&2986 &3429 \\ \hline 
$0^{++}$  &2074& 2536&2986 &3429 \\ \hline 
$1^{- -}$  &2149 & 2647& 3125& 3592\\ \hline  
$1^{+ -}$  &2149 & 2647& 3125& 3592\\ \hline  
$1^{+ +}$ &2149 & 2647& 3125& 3592\\ \hline 
\end{tabular}  \\[0.3cm]
exotic hybrids \\
\begin{tabular} {|c |c| c| c |c |c |c|}
\hline
 & n=0 & n=1 & n=2 & n=3 \\ \hline
$0^{+ -}$  &2074& 2536&2986 &3429 \\ \hline  
$0^{- -}$   &2074& 2536&2986 &3429 \\ \hline  
$1^{- +}$  &2149& 2647&3125 &  3592\\ \hline 
\end{tabular}  
\caption{The hybrid masses obtained
from Eqs. (\ref{scalar}) and (\ref{vector}) with the values of the conformal masses shown in  Eqs. (\ref{mesonicfields})-(\ref{exoticfields}).}
\label{Hmasses}
\end{center}
\end{table}

Eqs. (\ref{scalar}) and (\ref{vector}) have {been numerically solved } and
{the corresponding results are displayed in } Table \ref{Hmasses}. 

Since the GWS is supposed to be a  faithful $1/N_C$ leading order representation of $QCD$, we cannot consider e.g.,  the gluon a mass as {addressed for other} conventional model calculations.

\section{Lattice QCD and model calculations for light hybrids}
\label{lightlattice}
Due to the lack of experimental data, in order to test the present model, we will compare the GSW predictions with those of other approaches. In particular, 
the spectra of these hybrids have been studied by Lattice QCD (LQCD) \cite{Luo:2002rz,Dudek:2011bn,Bali:2003jq}, sum rule methods \cite{Reinders:1984sr,Kisslinger:1995yw,Burden:1996nh,Burden:2002ps} and different models \cite{Chanowitz:1982qj,Barnes:1982tx,Isgur:1984bm,Barnes:1995hc,Ishida:1991mx,General:2006ed}. 
Following the lines of Refs. \cite{Ketzer:2012vn,Meyer:2015eta,Luo:2002rz}, we report in
 Table \ref{Hmasses1} the main results of Refs. \cite{Luo:2002rz,Dudek:2011bn,Bali:2003jq,Reinders:1984sr,Kisslinger:1995yw,Burden:1996nh,Burden:2002ps,Chanowitz:1982qj,Barnes:1982tx,Isgur:1984bm,Barnes:1995hc,Ishida:1991mx,General:2006ed} compared with the predictions of the GSW model already 
 shown in 
  Table \ref{Hmasses}.
  We will specify the name of the model in the Table as used in the cited references. The masses will be given in GeV unity.

\begin{table} [htb]
\begin{center}
  Hybrid masses as in ref. \cite{Ketzer:2012vn}\\
\begin{tabular} {|c |c| c| c |c |c |c|c|}
\hline
 & Bag \cite{Chanowitz:1982qj,Barnes:1982tx}  & Flux Tube \cite{Isgur:1984bm,Barnes:1995hc} & Constituent Gluon \cite{Ishida:1991mx,General:2006ed}& LQCD ($m_\pi=396$ MeV) \cite{Dudek:2011bn}& GSW \\ \hline
$0^{-+}$ &1.3& 1.7-1.9&1.8-2.2 &2.1& $2.074\pm 0.028$ \\ \hline 
$0^{++}$  &$>$1.9& - &1.3-2.2 &$>$2.4&$2.074\pm 0.028$ \\ \hline 
$1^{- -}$  &1.7& 1.7-1.9&1.8-2.2& 2.3& $2.149 \pm 0.017$ \\ \hline 
$1^{+ -}$  &$>$1.9& 1.7-1.9&1.8-2.1& $>$2.4& $2.149 \pm 0.017$\\ \hline 
$1^{+ +}$ &$>$1.9 & 1.7-1.9& 1.3-2.2& $>$2.4&$2.149 \pm 0.017$ \\ \hline 
$0^{+ -}$ &- & 1.7-1.9& - &$>$2.4 &$2.074\pm 0.028$\\ \hline 
$0^{- -}$  &- & - &1.8-2.3 &  - &$2.074\pm 0.028$\\ \hline 
$1^{- +}$  &1.5& 1.7-1.9&1.8-2.2 &  2.0&$2.149 \pm 0.017$\\ \hline 
\end{tabular}  \\[0.3cm]
Hybrid masses as in refs. \cite{Meyer:2015eta,Luo:2002rz}\\ 
\begin{tabular} {|c |c| c| c |c |c |c|}
\hline
  & Sum Rule \cite{Reinders:1984sr,Kisslinger:1995yw} & Bethe-Salpeter \cite{Burden:1996nh,Burden:2002ps}& LQCD(renormalon) \cite{Bali:2003jq}& LQCD(anysotropic) \cite{Luo:2002rz}& GSW \\ \hline
$0^{++}$  & - &-&1.98&-&$2.074\pm 0.028$ \\ \hline 
$1^{- -}$  & - & -  &0.87 &-&$2.149 \pm 0.017$ \\ \hline 
$1^{+ -}$  & - &-  &1.25 &-&$2.149 \pm 0.017$\\ \hline 
$0^{+ -}$ & - & 1.082&-&-&$2.074\pm 0.028$ \\ \hline 
$0^{- -}$ & - &1.319  &- &-&$2.074\pm 0.028$ \\ \hline 
$1^{- +}$  & 1.3-1.8&1.439& 2.15&2.013 &$2.149 \pm 0.017$ \\ \hline 
\end{tabular}  
\caption{We show the masses obtained for the hybrid hadrons, whose  quantum numbers are defined above in Eqs. (\ref{mesonicfields})-(\ref{exoticfields}) and in refs. \cite{Ketzer:2012vn,Meyer:2015eta,Luo:2002rz,Dudek:2011bn,Reinders:1984sr,Kisslinger:1995yw, Chanowitz:1982qj,Barnes:1982tx,Isgur:1984bm,Barnes:1995hc,Ishida:1991mx,General:2006ed,Burden:1996nh,Burden:2002ps}.   The theoretical errors in the GSW predictions are due to the uncertainty on the $\alpha$ parameter. Masses given in GeV unity.}
\label{Hmasses1}
\end{center}
\end{table}
The GSW model calculations leads to values of the hybrid masses within the range
(2.07-2.15 GeV), similar to other model calculations \cite{Chanowitz:1982qj,Barnes:1982tx,Isgur:1984bm,Barnes:1995hc,Ishida:1991mx,General:2006ed,Burden:1996nh,Burden:2002ps}. 
However, as one can notice,  the masses of the lightest particles  differ between the predictions of mentioned collaborations. For the much sought exotic $1^{-+}$, the corresponding mass, obtained from the GSW model, is similar to that of the three LQCD calculations shown \cite{Luo:2002rz,Dudek:2011bn,Bali:2003jq}  but too high if compared with the Bethe-Salpeter approach~\cite{Burden:1996nh,Burden:2002ps}. {We also agree quite well with the LQCD\cite{Dudek:2011bn} and LQCD(renormalon) calculation of Ref. \cite{Bali:2003jq} in three states. However, our $0^{+-}$ and $0^{--}$ appear to be very high compared to the Bethe-Salpeter approach~\cite{Burden:1996nh,Burden:2002ps}, but reasonable compared with the Constitutent Gluon~\cite{Ishida:1991mx,General:2006ed} and low compared to LQCD~\cite{Dudek:2011bn}}.
{To summarize, we report that the $0^{-+}$ state is well reproduced while,  the other states are underestimated except the $1^{-+}$ case. Remarkably, these calculations, which do not involve any free parameter, are in line with those of other models or LQCD. In particular, we predict that scalar hybrids are lighter than the vector ones, as addressed
 by lattice data, except for the $1^{-+}$. However, as one might notice, the Lattice calculations clear indicate that only few states can be degenerate, for example, the mass of the $0^{-+}$ is lower than that of the $0^{++}$. In order to remove this degeneracy, in the next section we propose different possible modifications of the GSW model.}{Let us remark that, in the next section, only some possible simple  extensions of the GSW model will be discussed in order to highlight  how the present approach can be improved without the inclusion of further free parameters. Such a choice is essential to preserve the relevant predict power of the present model.
 It is also worth  stressing that the holographic approach represents a $1/N_c$ leading order calculation, which is good to determine hybrids with masses around 2 GeV.  }

\section{Beyond the GSW model}
\label{modi}
{Our aim here is to  propose an extended scenario to reproduce the hierarchy of the hybrid masses predicted by the lattice QCD and model calculations. The improvement that we present is  based on approaches already discussed in several investigations involving holographic models \cite{MartinContreras:2018nhv,Vega:2010ne,Forkel:2010gu}. All these scenarios do not require any additional free parameter.
As one might notice, since the present experimental and lattice scenarios are not well constrained, 
the main purpose of this section is to show that the GSW can be considered as a solid baseline for any future calculation or comparisons. In fact, it has been proved that the present approach is able to reproduce the almost linear trajectory of the glueball masses \cite{Rinaldi:2017wdn,Rinaldi:2018yhf,Rinaldi:2020ssz}, the spectra of light and heavy mesons (scalars, pseudo-scalars, vectors and axial vectors) \cite{Rinaldi:2020ssz,Rinaldi:2021dxh} with only two parameters. Let us also point out that the spirit of the modifications we are proposing is to parametrize the possible peculiar dynamics underlying the hybrid structure. In fact, we propose to keep the same 
 wrap factor in the metric in Eq. (\ref{metric5}), the profile function of the dilaton Eq. (\ref{prefactor}) and the parameters characterzing the model. Therefore, in order to describe hadrons with different dynamics from that of glueballs and regular mesons, some adjustment are needed. In particular, 
  to improve the model we consider that 
the field interpolators might lead to anomalous dimensions that might affect the conformal mass~\cite{MartinContreras:2018nhv,Vega:2010ne,Forkel:2010gu}:}

\begin{equation}
M_5^2R^2 = (\Delta + \Delta_p - p)(\Delta +\Delta_p +p -4).
\label{M5Am}
\end{equation}
{
However, there is no direct correspondence between the anomalous dimensions in QCD and the corresponding holomorphic anomalous dimensions. Let us discuss in what follows two modifications of the GSW associated with proposal $\Delta_p$.}

{
\subsection{First modification}
The first modification consists in introducing the anomalous dimension  $\Delta_p$ that leads to distinguish scalar and vector fields from the pseudoscalar and the axialvector ones.
Such a strategy  was quite successful in the study of regular mesons. In this case, one assigns $\Delta_p = -1$ for states whose field operator definition involves the $\gamma_5$ matrix~\cite{MartinContreras:2018nhv}. Using this input and Eq (\ref{M5Am})
one obtains the following conformal masses for each field interpolator of the mesonic type,} 

\begin{align}
 0^{- +} \; \;\;\;\;\,\;S& =  \bar{\Psi} \gamma^i\lambda^a \Psi B_i^a \qquad \qquad \;(\Delta = 5, \Delta_p  =\;\;0, p=0, M_5^2R^2=\;5), \nonumber \\
 0^{+ +} \; \;\;\;\:\:S^{\prime} &=  \varepsilon_{i j k} \bar{\Psi}  \sigma^{i j} \lambda^a \Psi  B^{k a} \qquad (\Delta = 5, \Delta_p  =\;\; 0, p=0, M_5^2R^2=\;5), \nonumber \\
 1^{- -} \; \;\;\, \;\;V_i&=  \bar{\Psi} \gamma_5  \lambda^a \Psi B_i^a \qquad \qquad \;(\Delta = 5,\Delta_p  = -1 , p=1, M_5^2R^2=\;3),\nonumber \\
  1^{+-}\;\;\;\;\; A^\prime_i &=\varepsilon_{i j k} \bar{\Psi} \gamma_5\gamma^j \lambda^a \Psi B^{k a}\; \quad (\Delta = 5, \Delta_p  = -1, p=1, M_5^2R^2=\; 3), \nonumber \\
 1^{+ +} \; \;\;\;\;A_i &=  \varepsilon_{i j k} \bar{\Psi} \gamma^j \lambda^a \Psi E^{k a} \qquad \;(\Delta = 5, \Delta_p  =\;\; 0, p=1, M_5^2R^2=\; \,8). \nonumber \\
\label{mesonicfields2}
  \end{align}
For the non-mesonic type, we get:

\begin{align}
 0^{+ -} \; \;\;\;\;\,\, \;\Sigma& =  \bar{\Psi} \gamma_5 \gamma^i \lambda^a \Psi B_i^a \qquad \qquad (\Delta = 5, \Delta_p  =\; -1, p=0, M_5^2R^2=\;0), \nonumber \\
 0^{- -} \; \;\;\,\, \;\;\Sigma^\prime&=  \bar{\Psi} \gamma_5  \lambda^a \Psi E_i^a \qquad \qquad  \quad (\Delta = 5,\Delta_p  =\; -1 , p=0, M_5^2R^2=\;0),\nonumber \\
 1^{- +} \; \;\;\;\;W_i &=  \varepsilon_{i j k} \bar{\Psi} \gamma^j \lambda^a \Psi B^{k a} \qquad \quad (\Delta = 5, \Delta_p  =\;\;\; 0, p=1, M_5^2R^2=\;8), \nonumber \\
 1^{- +} \; \, \;\; \, \,W_i^{\prime} &=  \bar{\Psi} \gamma_0\lambda^a \Psi  E_i^a \qquad \qquad \quad (\Delta = 5, \Delta_p \;\; \;\;= 0, p=1, M_5^2R^2=\;\,8). \nonumber \\
\label{exoticfields2}
\end{align}

{
The resulting spectrum is displayed in Tab. \ref{Hadronmasses3} in the columns labeled GSWm1.
As one can see, now the degeneracy between the first scalars ($0^{-+}$ and $0^{++}$) and the second ones ($0^{+-}$ and $0^{--}$) is removed, as that for the two vectors $1^{--}$ and $1^{+-}$  compared to $1^{++}$ and $1^{-+}$. However, in this case, except for the $1^{-+}$  we predict vector hybrids lighter than the scalar ones. Moreover, the masses of the $1^{--}$ and the $1^{+-}$ states largely underestimate the lattice predictions. 
We conclude that this simple modification based on the phenomenology of regular mesons does not lead to a significant improvement to the GSW model. On the contrary, the hierarchy between the scalar and vector mesons is not reproduced.}

\subsection{Second modification}
{ By
following the line of the procedure described in the previous section, but trying to take into account differences between regular and hybrid mesons,  we propose to introduce the anomalous dimension to eliminate the degeneracy between states with different parity, as proposed in Refs. \cite{MartinContreras:2018nhv,Vega:2010ne}. However, we assume that the lowest states, the $0^{-+}$ and the $1^{--}$ correspond to $\Delta_p=0$, while, those with opposite parity are associated to $\Delta_p=1$.
This strategy is almost equivalent to that discussed in Ref. \cite{MartinContreras:2018nhv,Vega:2010ne}, where there is an exchange of one unity between states with different parity. This modification is necessary just to reproduce the hierarchy of the masses displayed in Tab. \ref{Hadronmasses3}. Now, the conformal masses for the mesonic hybrids read:}

\begin{align}
 0^{- +} \; \;\;\;\;\,\;S& =  \bar{\Psi} \gamma^i\lambda^a \Psi B_i^a \qquad \qquad \;(\Delta = 5, \Delta_p  =\;\;0, p=0, M_5^2R^2=\;5), \nonumber \\
 0^{+ +} \; \;\;\;\:\:S^{\prime} &=  \varepsilon_{i j k} \bar{\Psi}  \sigma^{i j} \lambda^a \Psi  B^{k a} \qquad (\Delta = 5, \Delta_p  =\;\; 1, p=0, M_5^2R^2=\;12), \nonumber \\
 1^{- -} \; \;\;\, \;\;V_i&=  \bar{\Psi} \gamma_5  \lambda^a \Psi B_i^a \qquad \qquad \;(\Delta = 5,\Delta_p  = 0 , p=1, M_5^2R^2=\;8),\nonumber \\
  1^{+-}\;\;\;\;\; A^\prime_i &=\varepsilon_{i j k} \bar{\Psi} \gamma_5\gamma^j \lambda^a \Psi B^{k a}\; \quad (\Delta = 5, \Delta_p  = 1, p=1, M_5^2R^2=\; 15), \nonumber \\
 1^{+ +} \; \;\;\;\;A_i &=  \varepsilon_{i j k} \bar{\Psi} \gamma^j \lambda^a \Psi E^{k a} \qquad \;(\Delta = 5, \Delta_p  =\;\; 1, p=1, M_5^2R^2=\; \,15). \nonumber \\
\label{mesonicfields3}
  \end{align}
For the non-mesonic hybrids they become,

\begin{align}
 0^{+ -} \; \;\;\;\;\,\, \;\Sigma& =  \bar{\Psi} \gamma_5 \gamma^i \lambda^a \Psi B_i^a \qquad \qquad (\Delta = 5, \Delta_p  =\; 1, p=0, M_5^2R^2=\;12), \nonumber \\
 0^{- -} \; \;\;\,\, \;\;\Sigma^\prime&=  \bar{\Psi} \gamma_5  \lambda^a \Psi E_i^a \qquad \qquad  \quad (\Delta = 5,\Delta_p  =\; 0 , p=0, M_5^2R^2=\;5),\nonumber \\
 1^{- +} \; \;\;\;\;W_i &=  \varepsilon_{i j k} \bar{\Psi} \gamma^j \lambda^a \Psi B^{k a} \qquad \quad (\Delta = 5, \Delta_p  =\;\;\; 0, p=1, M_5^2R^2=\;8), \nonumber \\
 1^{- +} \; \, \;\; \, \,W_i^{\prime} &=  \bar{\Psi} \gamma_0\lambda^a \Psi  E_i^a \qquad \qquad \quad (\Delta = 5, \Delta_p \;\; \;\;= 0, p=1, M_5^2R^2=\;\,8). \nonumber \\
\label{exoticfields3}
\end{align}
{Let us refer to this modification as GSWm2. As one can see in Tab. \ref{Hadronmasses3}, the present free parameter approach, is in agreement with the predictions of the recent lattice calculations \cite{Dudek:2011bn}  except for the state $1^{--}$ which is slightly  underestimated. Of course, this is just an example of how a simple modification could lead to a good description of the present knowledge of hybrid states. Therefore, let us stress again that the GSW model in general can be also used to  provide useful predictions for the Physics of exotic hadrons.
For example, we might use the GSW model calculations to try to identify which states, among those addressed in the Particle Data Group, could be considered as hybrid candidates.}

\begin{table} [htb]
\begin{tabular} {|c |c| c| c |c |c |c|}
\hline
 & LQCD(renormalon) \cite{Bali:2003jq}  & QCD(anysotropic) \cite{Luo:2002rz} & LQCD ($m_\pi=396$ MeV) \cite{Dudek:2011bn}& GSW&GSWm1&GSWm2 \\ \hline
$0^{-+}$ &- & -&2.1& $2.074\pm 0.028$&$2.074\pm 0.028$&$2.074\pm 0.028$ \\ \hline 
$0^{++}$ &1.98 & - &$>$2.4&$2.074\pm 0.028$&$2.074\pm 0.028$&$2.694 \pm0.021$ \\ \hline 
$1^{- -}$  &0.87&-& 2.3& $2.149 \pm 0.017$& $1.562\pm 0.023$&$2.149 \pm 0.017$ \\ \hline 
$1^{+ -}$  &1.25& -&$>$2.4& $2.149 \pm 0.017$& $1.562\pm 0.023$&$2.747 \pm 0.013$\\ \hline 
$1^{+ +}$ &- & -&  $>$2.4&$2.149 \pm 0.017$&$2.149 \pm 0.017$&$2.747 \pm 0.013$ \\ \hline 
$0^{+ -}$ &- & -&$>$2.4 &$2.074\pm 0.028$&$1.411 \pm 0.052$&$2.694 \pm0.021$\\ \hline 
$0^{- -}$  &- & -  &  - &$2.074\pm 0.028$&$1.411 \pm 0.052$&$2.074\pm 0.028$\\ \hline 
$1^{- +}$  &2.15& 2.013&  2.0&$2.149 \pm 0.017$&$2.149 \pm 0.017$&$2.149 \pm 0.017$\\ \hline 
\end{tabular}  \\[0.3cm]
\caption{We show the masses obtained for the hybrid hadrons, whose  quantum numbers are defined above in Eqs. (\ref{mesonicfields})-(\ref{exoticfields}) and in refs. \cite{Bali:2003jq,Luo:2002rz,Dudek:2011bn}. {We compare the lattice calculations with the GSW predictions together with its modifications.   The theoretical errors in the GSW predictions are due to the uncertainty on the $\alpha$ parameter.}}
\label{Hadronmasses3}
\end{table}

\section{Phenomenological analysis for light hybrids}
\label{lightpheno}
Let us discuss our results in light of the experimental data for the spectra \cite{Workman:2022ynf}. {We recall that
the hybrids states $0^{-+}, 0^{++},1^{--},1^{++}$ have the same quantum numbers as regular mesons, hence} 
{one cannot directly consider the latter as hybrid states. However, we will proceed to compare the spectra of heavy mesons, reported  in Ref. \cite{Workman:2022ynf}, with our predictions for hybrids assuming that the model will guide us to identify which kind of these states might be hybrids. The corresponding masses are displayed in   }
 in Table \ref{PDGmasses1}.
{In the table, we also report the results for both the ground  (GSW$^{n=0}$) and the first  excited (GSW$^{n=1}$) states obtained from the GSW model. Since Lattice QCD calculations indicate the necessity of an  improvement of the present approach, we also include the masses predicted by the simple modification GSWm2 discussed in the previous section. 
From the comparisons between the theoretical predictions and the experimental data, one might conclude that from the point of view of the model, the following states could be hybrids:}

 \begin{itemize}
     \item[$0^{-+}$:] the $\pi(2070)$ and $\eta(2100)$ could be consistent with the ground state of a hybrid;
     \item[$0^{++}$:] the $f_0(2060)$ could be the ground state, while the $X(2540)$ could be a first excited state. If future improvements of lattice calculations will confirm the need for the modification of the GSW model, the $X(2540)$ could be also consistent with the ground state of the corresponding hybrid;

     \item[$1^{--}$:]  the $\phi(2170)$ and $\omega(2205)^*$ both have masses consistent with the hybrid ground state;
     \item[$1^{+-}$:] the $h_1(1965)^*$ could be a hybrid ground state;
     \item[$1^{++}$:] the $a_1(2095)$ could be a ground state hybrid.
     
 \end{itemize}
{ Let us remark that such a strategy is motivated by the  predictive power of this model, as reflected in Refs. \cite{Rinaldi:2017wdn,Rinaldi:2021dxh,Rinaldi:2022dyh}. Hence, we might infer that the model provides a reasonable depiction of certain aspects of QCD, and the utilization of its predictions to identify states potentially attributed to hybrids is well justified.}

\begin{table} [htb]
{\color{black}
\begin{center}
\begin{tabular} {|c |c| c| c |c | c | c | c | c | c |c|c|}
\hline
  &   $ \pi(1800)$ & $\eta(2010)^*$ & $\pi(2070)^*$ &$\eta(2100)^*$ &$\eta(2190)^*$&$\eta(2320)^*$&$\pi(2360)^*$ &GSW$^{n=0}$ &GSW$^{n=1}$ &GSW$_{m2}^{n=0}$ &GSW$_{m2}^{n=1}$  \\ \hline
$0^{-+}$ & $1800^{+9}_{-10}$ & $2010^{+35}_{-60}$ & $2070\pm35 $ & $ 2050^{+105}_{-50} $& $2190 \pm 50$ & $2320\pm 15$& $2360 \pm 25$ & $2074$ & $2536$ & $2074$ & $2536$  \\ \hline 
&   $f_0(1710)$ & $a_0(2020)^*$& $f_0(2060)^*$ &$X(2540)^*$& &&&GSW$^{n=0}$ &GSW$^{n=1}$ &GSW$_{m2}^{n=0}$ &GSW$_{m2}^{n=1}$ \\ \hline
$0^{++}$ & $1740^{+8}_{-7}$  & $ 2025 \pm 30$ & $ 2060 \pm 10$ & $ 2540^{+52}_{-28} $ & &&&2074&2536 & 2694 & 3179 \\ \hline 
&$\omega(1420)$&$\rho(1450)$& $\omega(1650)$& $\phi(1680)$&$\rho(1700)$&&&GSW$^{n=0}$ &GSW$^{n=1}$ &GSW$_{m2}^{n=0}$ &GSW$_{m2}^{n=1}$\\ \hline
$1^{--}$  & $ 1410 \pm 60$ & $ 1465 \pm 20$ &  $ 1670 \pm 30$  & $ 1680 \pm20 $ & $ 1720 \pm 20$ &&& 2149 & 2647 & 2149 & 2647 \\ \hline
&$\omega(1960)^*$&$\phi(2170)$&$\omega(2205)^*$&$\rho(2270)^*$&$\omega(2290)^*$&$\omega(2330)^*$&&GSW$^{n=0}$ &GSW$^{n=1}$ &GSW$_{m2}^{n=0}$ &GSW$_{m2}^{n=1}$ \\ \hline
$1^{--}$ &$ 1960 \pm 25$ & $ 2162 \pm 7$ & $ 2205 \pm 30$ & $ 2270 \pm 45$ & $ 2290 \pm 20$ & $2330 \pm 30$ &&2149&2647 &2149&2647\\ \hline 
& $h_1(1170)$ & $h_1(1415)$ & $ h_1(1595) $ & $ b_1(1960)^*$ & $h_1(1965)^*$& $ h_1(2215)^*$ & $b_1(2240)^*$&GSW$^{n=0}$ &GSW$^{n=1}$ &GSW$_{m2}^{n=0}$ &GSW$_{m2}^{n=1}$ \\ \hline
$1^{+-}$ & $1166 \pm 8$  &  $ 1416 \pm 8$ & $1594 ^{+25}_{-75}$ &  $ 1960 \pm 35$  & $ 1965 \pm 45$  & $2215 \pm 40$ & $ 2240\pm 35 $ &2149& 2647 & 2746 & 3251 \\ \hline 
& $a_1(1930)^*$ &$ f_1(1970)^*$ & $a_1(2095)^*$& $a_1(2270)^*$& $f_1(2310)^*$&&&GSW$^{n=0}$ &GSW$^{n=1}$&GSW$_{m2}^{n=0}$ &GSW$_{m2}^{n=1}$  \\ \hline
$1^{++}$  & $ 1930^{+30}_{-70} $ & $ 1971 \pm 15  $& $ 2096 \pm 138$ & $ 2270^{+55}_{-40} $ & $ 2310 \pm 60 $ &&& 2149& 2647 & 2746 & 3251 \\ \hline 
\end{tabular}  
\end{center}
\caption{{ We show the masses of the particles in the PDG whose quantum numbers correspond to mesons~\cite{Workman:2022ynf} and compare them with our calculated values for the hybrids for the first and the second modes. It must be noted that those particles marked with $^*$ are presented in PDG outside the summary table.}}
\label{PDGmasses1}
}
\end{table}

In Table \ref{PDGmasses1}  we observe that the masses of the candidate particles, many of which have not been considered or discovered, fall within the range of the lowest mode of our calculation in all cases, and some even fall within the range of the second mode.
Upon this analysis, it is clear that finding a pure hybrid state will be challenging, as they are likely to be mixed with mesons. In fact,
it is worth to notice  that the widths of these states are very large and therefore one might suspect that mixing of states could occur.
{In future investigations, we will consider applying the same strategy adopted in Ref. \cite{Rinaldi:2018yhf} where the GSW model has been used to establish the mixing condition between glueballs and meson states.}

More interesting are the other quantum numbers which do not correspond to known mesons $0^{+-}$, $0^{--}$ and $1^{-+}$. No particle appears in the PDG tables for the first two but, for the $1^{-+}$ we have probably one candidate, the $\pi_1 (2015)$ (see Tab. \ref{PDGmasses2}). 
In this case, as one can see, the reported mass is in agreement with the ground state predicted by the GSW model within the experimental error.
Let's conclude by noting that when considering the masses in our calculations of the $0^{+-}$ and $0^{--}$ states, they should be investigated. However, in this case, special decay properties need to be examined for a distinctive characterization.

\begin{table} [htb]
{\color{black}
\begin{center}
\begin{tabular} {|c |c| c| c |c | c | c|c| }
\hline
& &  &  & GSW$^{n=0}$ &GSW$^{n=1}$&GSW$_{m2}^{n=0}$ &GSW$_{m2}^{n=1}$\\ \hline
 $0^{+-}$ &  & &  &2074&2536 &2694 &3179  \\ \hline 
 && & & GSW$^{n=0}$ &GSW$^{n=1}$&GSW$_{m2}^{n=0}$ &GSW$_{m2}^{n=1}$ \\ \hline
$0^{--}$  && &   &2074&2536 &2074&2536 \\ \hline 
&$ \pi_1(1400)$ & $\pi_1(1600)$ & $\pi_1(2015)$& GSW$^{n=0}$ &GSW$^{n=1}$&GSW$_{m2}^{n=0}$ &GSW$_{m2}^{n=1}$\\ \hline
$1^{-+}$  & $ 1354 \pm 25$ & $ 1661^{+15}_{-11}$   & $ 2001 \pm 122 $  & 2149&2647&2149&2647 \\ \hline 
\end{tabular}  
\caption{{We show the masses of the particles in the PDG for the quantum numbers that do not correspond to mesons. The $\pi_1 (2015)$ has been omitted from the particle table~\cite{Workman:2022ynf}.}}
\label{PDGmasses2}
\end{center}
}
\end{table}

\section{The spectrum of the non light hybrids}
\label{heavyhybrid}

Since the holographic approach here adopted relies on conformal symmetry, predictions can be realistic once the chiral symmetry of QCD is restored. Hence, the proposed model does not contain any dependence on the flavor of the constituent quarks of the hadrons. Nevertheless, further modifications of the approach can be taken into account to reproduce the masses of heavy hadrons \cite{Rinaldi:2020ssz,Rinaldi:2021dxh}. In particular,
we apply the approach introduced in section \ref{lighthybrid}  to $s$,$c$ and $b$ quark-antiquark pairs. For this purpose, we follow the prescription addressed in Refs. \cite{Rinaldi:2020ssz,Rinaldi:2021dxh}, namely to add a constant to the mass of the light mesons

\begin{equation}
M_{heavy, \;n} = M_{light,\; n} + C.
\end{equation}
Let us report the values of the constant $C$ corresponding to the considered quark flavors \cite{Rinaldi:2020ssz,Rinaldi:2021dxh}:
$C_c=2400$MeV ,$C_b=8700$ MeV. We add  here also $C_s$ associated to strangeness which we has not studied before, $C_s=300$ MeV.
{We then add the above constants to the mass spectra predicted by the GSW model previously calculated, for ground and excited states. }
We show in Tables \ref{HHmasses0}-\ref{HHmasses01} the results obtained by performing this operation. We also apply the present procedure to the predictions  obtained also for  $\Delta_p=1$, i.e., the second modification.
In the next sections, we compare the results of this analysis with the predictions of other quark models and lattice QCD calculations.

\begin{table} [htb]
\begin{center}
 mesonic  hybrids ($\Delta_p=0$) \\
\begin{tabular} {|c |c| c| c |c |c |c| c| c| c| c| c | c |c |c |}
\hline
 &$\bar{s}s$ & n=0 & n=1 &$\bar{c}c$ & n=0 & n=1 &$\bar{b}b $& n=0&n=1\\ \hline
$0^{-+}$ &&2374& 2836&& 4474&4936&&10774&11236 \\ \hline 
$0^{++}$  &&2374& 2836&& 4474&4936&&10774&11236 \\ \hline 
$1^{- -}$  &&2449& 2947&&4549& 5074&& 10849& 11347\\ \hline 
$1^{+ -}$ & &2449& 2947&&4549& 5074&& 10849& 11347\\ \hline 
$1^{+ +}$& & 2449& 2947&&4549&5074&&10849&11347\\ \hline 
\end{tabular}  \\[0.3cm]
exotic hybrids ($\Delta_p=0$) \\
\begin{tabular} {|c |c| c| c |c |c |c| c| c| c| c|}
\hline
 &$\bar{s}s$  & n=0 & n=1 &$\bar{c}c$ & n=0 & n=1 &$\bar{b}b$ & n=0&n=1\\ \hline
$0^{+ -}$ &&2374& 2836&&  4474& 4936&& 10774&11236 \\ \hline 
$0^{- -}$  &&2374& 2836&&  4474& 4936&& 10774&11236 \\ \hline 
$1^{- +}$  &&2449& 2947&& 4549& 5074&& 10849&11347\\ \hline 
\end{tabular}  
\caption{We show the masses obtained for the heavy hybrid interpolating fields defined above, Eqs. (\ref{mesonicfields})-(\ref{exoticfields}) having fixed all our parameters with the scalar hadrons in Ref.~\cite{Rinaldi:2021dxh}.}
\label{HHmasses0}
\end{center}
\end{table}

\begin{table} [htb]
\begin{center}
 mesonic  hybrids ($\Delta_p=1$) \\
\begin{tabular} {|c |c| c| c |c |c |c| c| c| c| c| c | c |c |c |}
\hline
 &$\bar{s}s$ & n=0 & n=1 &$\bar{c}c$ & n=0 & n=1 &$\bar{b}b $& n=0&n=1\\ \hline
$0^{-+}$ &&2374& 2836&& 4474&4936&&10774&11236 \\ \hline 
$0^{++}$  &&2994& 3479&& 5094&5579&&11394&11879 \\ \hline 
$1^{- -}$  &&2449& 2947&&4549& 5074&& 10849& 11347\\ \hline 
$1^{+ -}$ & &3047& 3551&&5147& 5651&& 11447& 11951\\ \hline 
$1^{+ +}$& & 3047& 3551&&5147&5651&&11447&11951\\ \hline 
\end{tabular}  \\[0.3cm]
exotic hybrids ($\Delta_p=1$) \\
\begin{tabular} {|c |c| c| c |c |c |c| c| c| c| c|}
\hline
 &$\bar{s}s$  & n=0 & n=1 &$\bar{c}c$ & n=0 & n=1 &$\bar{b}b$ & n=0&n=1\\ \hline
$0^{+ -}$ &&2994& 3479&&  5094& 5579&& 11394&11879 \\ \hline 
$0^{- -}$  &&2374& 2836&&  4474& 4936&& 10774&11236 \\ \hline 
$1^{- +}$  &&2449& 2947&& 4549& 5074&& 10849&11347\\ \hline 
\end{tabular}  
\caption{We show the masses obtained for the heavy hybrid interpolating fields defined above, Eqs. (\ref{mesonicfields3})-(\ref{exoticfields3}) having fixed all our parameters with the scalar hadrons in Ref.~\cite{Rinaldi:2021dxh}.}
\label{HHmasses01}
\end{center}
\end{table}

\section{Lattice QCD and model calculations for heavy hybrids}
\label{heavylattice}
There have been many calculations interested in the study of heavy hybrid hadrons. In Table \ref{HHmasses1} we show some of their results~\cite{Luo:2002rz,Ma:2020bex,Chen:2013zia,Oncala:2017hop}

\begin{table} [htb]
\begin{center}
  Heavy Hybrid masses \\
\begin{tabular} {|c |c| c| c |c |c |c |c |c |c | }
\hline
 &LQCD& LQCD&LQCD&SR & SR& NRQCD & NRQCD &NRQCD  & NRQCD \\ \hline 
 &&&&&&&&&\\
 &  ($\bar{s}s$) \cite{Ma:2020bex} & $(\bar{s}s$) \cite{Ma:2020bex}& ($\bar{c}{c}$) \cite{Luo:2002rz} &($\bar{c}{c}$) \cite{Chen:2013zia}&SR($\bar{b}{b}$) \cite{Chen:2013zia}&($\bar{c}{c}$)\cite{Oncala:2017hop}&($\bar{c}{c}$)\cite{Oncala:2017hop}&($\bar{b}{b}$) \cite{Oncala:2017hop}&($\bar{b}{b}$ \cite{Oncala:2017hop}) \\ \hline 
 &n=0 &n=1 &n=0 &n=0 &n=0 &n=0& n=1 &n=0& n=1  \\ \hline
$0^{-+}$& 1.7&- &-&3.61&9.68&4.011&4.355&10.690&10.885 \\ \hline 
$0^{++}$   & - &- &-& 5.34&11.20&4.486&4.920&11.011&11.299\\ \hline 
$1^{- -}$  &1.7&-&-& 3.36& 9.70&4.011&4.355&10.690&10.885\\ \hline 
$1^{+ -}$  & -&-&-&4.53&10.70&4.145&4.511&10.761&10.970\\ \hline 
$1^{+ +}$  & -&-&-&5.06&11.09&4.145& 4.511&10.761&10.970\\ \hline 
$0^{+ -}$  & -& - &-&4.09&10.17&4.145&4.511&10.761&10.970 \\ \hline 
$0^{- -}$  & - &  - &&5.51&11.48&-&-&-&-\\ \hline 
$1^{- +}$  &2.1-2.2 &  3.6 &4.369&3.70&9.79&4.011&4.355&10690&10885\\ \hline 
\end{tabular}  
\caption{We show the masses obtained for the heavy hybrid hadrons, whose quantum numbers are defined by Eqs. (\ref{mesonicfields})-(\ref{exoticfields}), in refs. \cite{Luo:2002rz,Ma:2020bex,Chen:2013zia,Oncala:2017hop}}
\label{HHmasses1}
\end{center}
\end{table}

Let us proceed by comparing these spectra with the outcomes of the GSW model (Tab. \ref{HHmasses0}) and its modification (Tabs. \ref{HHmasses01}).  
Let us start from the lattice evaluation from Refs. \cite{Luo:2002rz,Ma:2020bex}-
The predictions of the model overestimate the  $s \bar s$ ground states  $0^{-+}$ and $0^{++}$. However, our calculations almost agree with the $1^{-+}$ ground and excited states  A good agreement is also found for the ground state of the $1^{-+}$ for the $c \bar c$ hadron ~\cite{Ma:2020bex,Luo:2002rz}. We also compare our results with the model of Ref. \cite{Chen:2013zia} (SR). The main differences can be found in the $0^{-+}$, the $1^{--}$ and the $1^{-+}$ states. Let us conclude with the comparison of the GSW model calculations, with those of Ref. \cite{Oncala:2017hop}(NRQCD). Here one can notice that  our predictions o are in line with the outcome of Ref. \cite{Oncala:2017hop} for the $c \bar c$ and $b \bar b$ states for $n=0$. However, the excited states are overestimated.

\section{Phenomenological analysis for heavy hybrids}
\label{heavypheno}
{In this section we will proceed as before. We will use the GSW model to examine the data and determine if they correspond to any states that could potentially be hybrids. In Table \ref{PDGheavymasses} we present particle masses from the PDG \cite{Workman:2022ynf} with mesonic quantum numbers also shared with possible heavy hybrids, and we compare them with the first two modes of our calculation.}

In the $s \bar s$ case, the prediction of the GSW model overestimates, for example, the $\phi(2170)$ state. For the $c \bar c$ hadrons,  the $\chi_{c0}(4500)$, the $\Psi(4660)$ and the $Z_c(4430)$ states are possible candidates. Finally, for $b \bar b$ hadrons the $\chi_{b0}(2P)$, the $\Upsilon(10860)$, the $\Upsilon(11020)$ are close to the spectra predicted by the GSW model. The $Z_b(10650)$ and the $\chi_{b1}(3P)$ are overestimated but, the disagreement is not particularly big.

\begin{table} [htb]
\begin{center}
\begin{tabular} {|c |c| c| c |c | c | c | c | c | c |}
\hline
$\bar{s}s$ hybrids&$\phi(1020)$&$\phi(1680)$& $\phi(2170)$&&&&& GSW & GSW$_{m2}$\\ \hline
$1^{--}$  & $ 1019.46 \pm 0.02 $ & $ 1680 \pm 20$ &  $ 2162 \pm 7$ & &&&&2449& 2449  \\ \hline \hline
 $\bar{c}c$ hybrids &    $\eta_c$(1S) & $\eta_c$(2S) &&&&& & GSW & GSW$_{m2}$  \\ \hline
$0^{-+}$ & $2983.9 \pm 0.4$ & $3637.5 \pm 1.1$   && & & & &4474& 4474  \\ \hline 
&   $\chi_{c 0}$ (1P) & $\chi_{c 0}(3860)$& $\chi_{c0}(3915)$  & $\chi_{c0}(4500)$ & $\chi_{c0}(4700)$ & & &GSW & GSW$_{m2}$ \\ \hline
$0^{++}$ & $3862 ^{+66}_{-55}$ & $ 3414.7 \pm 0.3 $  & $ 3921.7 \pm 1.8$ & $ 4474 \pm 6$ & $ 4694^{+20}_{-7} $ & & & 4474& 5094 \\ \hline 
& $\Psi(3770)$ & $\Psi(4040)$ & $\Psi (4160)$& $\Psi (4230)$ & $\Psi(4360)$ & $\Psi(4415)$ & $\Psi(4660)$ & GSW & GSW$_{m2}$\\ \hline
$1^{--}$ &  $ 3773.7 \pm 0.4$  & $4039 \pm 1$ &  $ 4191 \pm 5$  & $ 4222.6 \pm 2.6 $ & $ 4372 \pm 9$  & $ 4421 \pm 4$ & $ 4630 \pm 6 $  & 4549 & 4549 \\ \hline
&$h_c$(1P) & $Z_c(3900)$ & $Z_c(4200)$ & $Z_c(4430)$ & & & &   GSW & GSW$_{m2}$\\ \hline
$1^{+-}$ & $ 3525.38 \pm 0.11$  & $3887.1 \pm 2.6$ & $ 4196^{+48}_{-42} $  & $ 4478^{+15}_{-18}$ & & & &4549& 5147\\ \hline 
& $\chi_{c 1}$ (1P) & $\chi_{c 1}(3872)$&$\chi_{c 1} (4140)$ & $\chi_{c 1}(4274)$&&&& GSW & GSW$_{m2}$ \\ \hline
$1^{++}$  & $ 3510.67 \pm  0.05 $ & $ 3871.65 \pm 0.06$  & $4146.5 \pm 3.0$ & $ 4286^{+8}_{.9}$ && && 4549&5147 \\ \hline \hline
$\bar{b}b$ hybrids  &    $\eta_b$(1S) & &&&&&  &GSW & GSW$_{m2}$  \\ \hline
$0^{-+}$ & $ 9398.7 \pm 2.0$ & && & & && 10774 & 10774 \\ \hline 
&   $\chi_{b 0}$ (1P) & $\chi_{b 0}(2P)$&& &&&& GSW & GSW$_{m2}$ \\ \hline
$0^{++}$ &$ 9859.44 \pm 0.73$  & $ 10232.5 \pm 0.9$ && &&&& 10774&11394 \\ \hline 
&$\Upsilon$(1S)&$\Upsilon$(2S)& $\Upsilon$(3S)& $\Upsilon$ (4S)&$\Upsilon(10860)$&$\Upsilon(11020)$&&GSW & GSW$_{m2}$\\ \hline
$1^{--}$  &$ 9460.30 \pm 0.26$ & $10023.26\pm 0.31$ & $  10355.2 \pm 0.31$  & $10579.4 \pm 1.2$ & $ 10885.2^{+2.6}_{1.6} $  &$ 11000 \pm 4$ & &10849& 10849  \\ \hline
&$h_b$(1P)&$h_b$(2P)&$Z_b(10610)$&$Z_b(10650)$&&&&  GSW & GSW$_{m2}$\\ \hline
$1^{+-}$ & $ 9899.3 \pm 0.8$ & $10259.8 \pm 1.6 $ & $ 10607.2 \pm 2.0$ & $ 10652.2 \pm 1.5$&&&& 10849& 11447\\ \hline 
& $\chi_{b 1}$ (1P) & $\chi_{b 1}$(2P)&$\chi_{b 1} $(3P)& &&&& GSW & GSW$_{m2}$ \\ \hline
$1^{++}$  & $9892.78 \pm 0.57$ & $10255.46 \pm 0.72$ & $10513.42\pm 0.94$ &&& &&10849&11447\\ \hline 
\end{tabular}  
\end{center}
\caption{{We show the masses of the particles in the PDG whose quantum numbers addressed in the present analysis but corresponding to mesons~\cite{Workman:2022ynf} and compare them with the results of the calculattions of the heavy hybrid spectra with the GSW model and its modification (GSW$_{m2}$).}}
\label{PDGheavymasses}
\end{table}

\section{Conclusions}
\label{conclusions}

{We have used the Graviton Soft Wall model, previously developed for conventional hadrons, to analyze hybrid hadrons. 
To this aim, we have computed the conformal masses by examining the potential minimal p-form field configuration that can be obtained from quark and gluon fields. 
These configurations define the spins and parities of the hybrids and determine the corresponding conformal masses, which in turn characterize the corresponding bound state equations in the fifth dimension.
It is important to highlight that our calculation is parameter free, as the two parameters employed in the approach have been determined by the scalar glueballs and mesons~\cite{Rinaldi:2020ssz,Rinaldi:2021dxh}.
Additionally, for the heavy hybrids, we have included the same parameters as those used for heavy mesons, which essentially represent the heavy quark masses.}
{The results obtained in our parameter free calculation has been compared with other model calculations. We have analyzed in detail the similarities and differences. We tend to agree with the lattice results and the NRQCD better than with the SR approach.}
{Moreover, we found out that the GSW model, due to the simplicity lowest order p-forms, leads to the same conformal mass for different states and therefore to mass degeneracies. Looking back at table~\ref{Hmasses} we see degeneracies between the $0^{-+}$, the $0^{++},$ the $0^{+-}$ and the $0^{--}$ hybrids, and the $1^{--}$, the $1^{+-}$, the $1^{++}$ and the  $1^{-+}$ hybrids. }
{In order to remove such a degeneracy, not predicted by recent lattice calculations, we took into account the effeccts of anomalous dimensions for some of these state and, hence, the corresponding degeneracies are eliminated, as can be seen in Table \ref{Hadronmasses3} for the two modifications studied. The remaining degeneracies correspond to underlying symmetries not accounted for in QCD. Therefore, additional adjustments to the naive GSW model should be pursued for more accurate predictions.}
{Finally, we have proposed a different approach to analyze  the PDG spectra. We assume our model's results are accurate mass values for hybrids and attempt to identify potential states in the data using these mass values, indicating the possible presence of hybrids in the spectra.
In this way we have predicted several possible hybrid states but, in many cases given the closeness to conventional mesons states, one expects strong mixing between mesons and hybrids.}

\section*{Acknowledgements}
MR thanks for the financial
support received under the STRONG-2020 project of the
European Union’s Horizon 2020 research and innovation
programme: Fund no 824093. VV was
supported by MCIN/AEI/10.13039/501100011033, European Regional Development Fund Grant No. PID2019-105439 GB-C21 and by GVA PROMETEO/2021/083.

\bibliographystyle{unsrt}

\bibliography{GSW.bib}

\end{document}